\documentclass[aps,nofootinbib,showpacs,twocolumn,floatfix]{revtex4}
\usepackage{epsf,amsfonts,amssymb,amsbsy}

\begin{document}
\title{Strong CP $\boldsymbol\theta$-problem in QCD and local gluon
condensate}
\author{V.V.Kiselev}
\email{kiselev@th1.ihep.su}
%\fax{}
\affiliation{Russian State Research Center "Institute for High
Energy Physics",%\\
Protvino, Moscow Region, 142281, Russia\\ Fax: +7-0967-744739}
%\date{}
\pacs{12.38.Aw, 14.80.Mz, 11.30.Er}
\begin{abstract}
We calculate the contribution of local gluon condensate into the
quadratic $\theta$-term of vacuum energy and discuss its
application to the study of axion mass.
\end{abstract}
\maketitle

%%%\hbadness=1500
\section{Introduction}

The strong CP problem reveals itself by the term of lagrangian
permitting the CP-violation in an explicit form
\begin{equation}
{\cal L}_{\theta} = - \theta\;  \frac{\alpha_s}{8\pi}
\;\epsilon_{\mu\nu\alpha\beta}\, G^{\mu\nu}_a (x)
G^{\alpha\beta}_a (x), \label{bare}
\end{equation}
where \textit{a priori} free quantity $\theta$ denotes the
effective parameter including the bare contribution as well as the
term caused by the matrix of quark masses $M$,
$$
\theta = \theta_{bare} - \mbox{arg}(\det M).
$$
The smallness of $\theta$ following from the experimental
observations, basically, by the constraints on the electric dipole
moment of neutron, looks rather accidental in QCD, unless we do
not involve some additional mechanisms. The most elegant
opportunity is the Peccei--Quinn symmetry \cite{PQ} introducing an
axion field coupled to the CP-odd product of gluon strength
tensors in (\ref{bare}). In that case, the quantity $\theta$ gains
the dynamical nature due to the axion $a$, so that $\theta\to
\theta+a/f_a$, where we have introduced the axion decay constant
$f_a$, and the problem is solved by an appropriate effective
potential for $\theta$, if the minimal energy is reached at
$\langle\theta\rangle=0$. Since the bare and quark mass terms in
$\theta$ lead to a global additive redefinition of axion field, we
can easily see that the positive mass squared of axion in the
effective action is a necessary condition in order to control the
stability, while the position of stable point should be at
$\theta=0$ to get the above solution of strong CP-problem.

The real value of axion mass depends on whether the axion is
coupled to the quark fields or not. The corresponding term of
fermion contribution is given by
\begin{equation}\label{fermion}
    {\cal L}_f = {\rm i} g_{n}\, a\,\bar q_n\gamma_5 q_n,
\end{equation}
where $n$ marks the quark flavor \cite{KSVZ}. There is a
possibility to introduce the coupling to the leptons, too,
\cite{DFSZ}. So, due to the fermion component the mixing of axion
with the pseudo-Goldstone bosons appearing under the spontaneous
breaking of chiral symmetry results in the following general
formula \cite{CKK} for the axion mass:
\begin{equation}\label{chir}
    m_a^2 = \frac{N_a^2}{f_a^2}\, \frac{\langle\bar q q\rangle K}
    {\langle\bar q q\rangle -K\textrm{Tr} M^{-1}},
\end{equation}
where $N_a$ is the color anomaly of Peccei--Quinn current, which
we put equal to 1, $\langle\bar q q\rangle $ denotes the quark
condensate\footnote{Further, one usually insets the relation
$(m_u+m_d)\langle\bar q q\rangle = - m_\pi^2\,f_\pi^2$ with
$\langle\bar q q\rangle = \langle\bar u u\rangle+\langle\bar d
d\rangle$ and $f_\pi \approx 93.3$ MeV.}, while $K$ is a
dimensional quantity ordinary represented by the instanton-induced
term in the effective potential. Therefore, the light quarks
coupled to the axion essentially change the situation, since they
suppress purely gluon contribution. Nevertheless, the
consideration of gluon term in the quadratic $\theta$-action
remains actual as the leading correction to the quark condensate,
if the axion is coupled to the quarks, or as the main term, if the
axion is coupled to the gluons, only.

In the present study we calculate the effective action quadratic
in $\theta$ due to the contribution of the local gluon condensate.
This contribution can be competitive with the term induced by the
instantons, since, first, as observed by the lattice simulations,
the instantons do not dominate in the determination of vacuum
properties in QCD because the vacuum is too warm, second,
modelling the gluon condensate by the dilute instanton gas is
problematic as the gas should consist of rather wide and dense
instantons in order to reach the observed magnitude of the gluon
condensate. In addition, the distribution over the instanton sizes
involves extra model-parameters. Thus, the gluon condensate
contribution itself is of interest.

\section{Calculations and results}

\begin{figure}[th]
\setlength{\unitlength}{0.9mm}
\begin{center}
\begin{picture}(100,45)
\put(5,0){\epsfxsize=85\unitlength \epsfbox{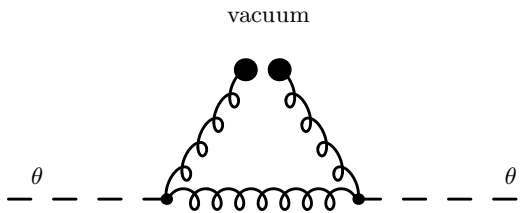}}
\put(10,15){$\theta$} \put(80,15){$\theta$} \put(39,39){vacuum}
\end{picture}
\end{center}

\vspace*{-1cm} \caption{The quadratic $\theta$-term caused by the
local gluon condensate.} \label{fglu}
\end{figure}

In the calculation of diagram shown in Fig. \ref{fglu} we take the
leading term caused by the local gluon condensate,
%\begin{widetext}
\begin{eqnarray}
&&\langle G^{\mu\nu}_a(x) G^{\mu'\nu'}_b(y)\rangle = \langle
G^{\mu\nu}_a(0) G^{\mu'\nu'}_b(0) \rangle\nonumber \\ &&
~~+y^{\alpha} \langle G^{\mu\nu}_a(0) \partial_{\alpha}
G^{\mu'\nu'}_b(0)\rangle + x^{\alpha}\langle
\partial_{\alpha}G^{\mu\nu}_a(0) G^{\mu'\nu'}_b(0)\rangle\nonumber
\\ && ~~+x^{\alpha}y^{\beta} \langle
\partial_{\alpha}G^{\mu\nu}_a(0)
\partial_{\beta}G^{\mu'\nu'}_b(0)\rangle+\ldots\nonumber \\ && ~~=
\langle G^{\mu\nu}_a(0) G^{\mu'\nu'}_b(0)\rangle +\mbox{higher
condensates}
\end{eqnarray}
%\end{widetext}
so that we can write down its contribution to the effective action
in the explicit form
\begin{eqnarray}
\textrm{i}\Gamma_{\theta\theta} &=& \lim_{k\to 0}\;
(-\textrm{i})^2 \left(\frac{\alpha_s}{8\pi}\right)^2
\epsilon_{\mu\nu\alpha\beta}\, \epsilon_{\mu'\nu'\alpha'\beta'}\;
\langle  G^{\mu\nu}_a G^{\mu'\nu'}_b \rangle {\scriptstyle \times}\,\nonumber\\
&& \textrm{i} k^{\alpha}\, (-\textrm{i} k^{\alpha'})\, \frac{-
\textrm{i} g^{\beta\beta'}}{k^2}\, \delta_{ab}\, F_{comb},
\end{eqnarray}
where the combinatoric factor $F_{comb} = 16$ is determined by the
permutations of gluon strength tensors and the evident relation
$\epsilon_{\mu\nu\alpha\beta} (\partial^{\alpha} A^{\beta}_a -
\partial^{\beta} A^{\alpha}_a) = 2 \epsilon_{\mu\nu\alpha\beta}
\partial^{\alpha} A^{\beta}_a $.

For the vacuum expectation value we evidently have the following
general expression conserving the color and Lorentz symmetries:
\begin{equation}
\langle  G^{\mu\nu}_a  G^{\mu'\nu'}_b  \rangle =
\frac{\delta_{ab}}{96}\;
(g^{\mu\mu'}g^{\nu\nu'}-g^{\mu\nu'}g^{\nu\mu'}) \; \langle
G^{\mu\nu}_d  G_{\mu\nu}^d \rangle . \label{real}
\end{equation}
Then, we find the gluon condensate part of quadratic term in the
effective potential for the global phase $\theta$
\begin{equation}
V_{\theta^2}[G^2] = \theta^2\; \left\langle
\frac{\alpha_s^2}{8\pi^2} G^{\mu\nu}_a  G_{\mu\nu}^a
\right\rangle,\label{glu}
\end{equation}
while the linear term due to the bare lagrangian in (\ref{bare}) does not
contribute into the vacuum energy, since the physical vacuum infers condition
(\ref{real}).

We can straightforwardly get the contribution given by the local
quark condensate into the quadratic term of axion in the way
analogous to the described calculation with the gluon condensate,
so that
\begin{equation}
V_{a^2}[\bar q q] = - a^2\;\sum_n \frac{|g_n|^2}{m_n}\,\langle
\bar q_n q_n\rangle,\label{quark}
\end{equation}
which can be further transformed after the substitution for the
universal coupling of the axion to the quarks \cite{CKK}
\begin{equation}\label{g_n}
    g_n = \frac{1}{f_a}\,\frac{1}{\mbox{Tr}\,M^{-1}},
\end{equation}
expressing a consistency for the definition of axion decay
constant through the calculation of axial anomaly and the absence
of mass mixing between the axion and neutral pseudo-Goldstone
mesons $\pi^0$ and $\eta$. Here, $M$ denotes the diagonal mass
matrix of light quarks ($u,\,d$). Then, we get
\begin{equation}
V_{a^2}[\bar q q] = - \frac{1}{2} \frac{a^2}{f_a^2}\; \langle \bar
q q\rangle\,\frac{m_um_d}{m_u+m_d},\label{quark2}
\end{equation}
which reproduces the result of \cite{CKK} in the case of quark
condensate dominance. We stress that the mixing of the axion with
the flavor-singlet meson leads to the quadratic axion potential
tending to zero in the limit of massless light quark in agreement
with the general formula given by (\ref{chir}) despite the
calculated term due to the local gluon condensate.

Indeed, comparing (\ref{quark2}) with (\ref{glu}) after the
substitution of $\theta\to a/f_a$ we could observe that the
contribution by the local gluon condensate dominates in the axion
mass. The picture is critically changed, if we take into account
the neutral singlet of SU(3) flavor group. This singlet is coupled
to the gluon anomaly, too, in the same manner as the axion with an
analogous constant $f_s$ instead of $f_a$, which produces the
mixing of masses for the axion and singlet \cite{CKK}. The
resulting effect has been described in the Introduction, and we
return to the quark condensate dominance in the axion mass up to
small corrections. The description of above mechanism is given in
\cite{CKK} in detail.

Nevertheless, we stress that expression (\ref{glu}) gives explicit
result in contrast to rather indefinite instanton-induced
potential.

Let us discuss the renormalization group (RG) properties of
results obtained. First, we can easily see that the contribution
by the quark condensate (\ref{quark2}) enters into the axion mass
in the RG invariant form, since the ratio of quark masses and the
product of quark mass with the condensate are invariant under the
action of RG. Second, the gluon condensate term due to (\ref{glu})
reveals the RG dependence with nonzero anomalous dimension in QCD.
However, this anomalous dimension is in the next $\alpha_s$-order
under study. Therefore, we have no any contradiction with the QCD
principles in the given accuracy over $\alpha_s$.

We do not make some numerical estimates, since they would be
rather qualitative under assumptions on the effective value of
$\alpha_s$.

Thus, we clarify the physical meaning of our calculations and
results.

\section{Conclusion}

We have calculated the contribution by the local gluon condensate
into the effective potential quadratic over the $\theta$
determining the CP-violating term of the gluon anomaly in QCD.
Such the effective potential gives the contribution into the axion
mass providing the solution of strong CP-problem in the
Peccei--Quinn mechanism. We expect the dominance of contribution
caused by the local gluon condensate in comparison with the
instanton-induced potential, which could be tested by the lattice
simulations.

The axion plays an essential role in the connection of particle
physics with the cosmology, in particular, as a dark matter
candidate \cite{Kh}. This fact makes the studies on its mass to be
rather significant. We see that the QCD dynamics in such the
investigations is of importance. We would like to point to some
recent interesting papers devoted to the problem under discussion:
solutions of strong CP-problem \cite{Gl,Ber}, on the Yang--Mills
vacuum \cite{Ch}.

This work is in part supported by the Russian Foundation for Basic
Research, grants 01-02-99315, 01-02-16585.

%\newpage

\end{document}